\documentclass[conference]{IEEEtran}
\IEEEoverridecommandlockouts
\usepackage{cite}
\usepackage{amsmath,amssymb,amsfonts}
\usepackage{algorithmic}
\usepackage{graphicx}
\usepackage{textcomp}
\usepackage{xcolor}
\usepackage{bigints}
\def\BibTeX{{\rm B\kern-.05em{\sc i\kern-.025em b}\kern-.08em
    T\kern-.1667em\lower.7ex\hbox{E}\kern-.125emX}}
\begin{document}
 \pdfoutput=1  
\title{Impact of Secondary User Interference \\ on Primary Network in Cognitive Radio Systems}

\author{\IEEEauthorblockN{Amit Kachroo and  Sabit Ekin}
\IEEEauthorblockA{School of Electrical and Computer Engineering\\
Oklahoma State University, Stillwater, OK, USA\\
Email: amit.kachroo, sabit.ekin\{@okstate.edu\} }
}

\maketitle

\begin{abstract}
Most of the research in cognitive radio field is primarily focused on finding and improving secondary user (SU) performance parameters such as bit error rate, outage probability and capacity etc. Less attention is being paid towards the other side of the network that is the primary network which is under interference from SU. Also, it is the primary user (PU) that decides upon the interference temperature constraint for power adaptation to maintain a certain level of quality of service while  providing access to SUs. However, given the random nature of wireless communication, interference temperature can be regulated dynamically to overcome the bottlenecks in entire network performance. In order to do so, we need to analyze the primary network carefully. This study tries to fill this gap by analytically finding the closed form theoretical expressions for signal to interference and noise ratio (SINR), mean SINR, instantaneous capacity, mean capacity and outage probability of PU, while taking peak transmit power adaptation at SU into picture. Furthermore, the expressions generated are validated with the simulation results and it is found that our theoretical derivations are in perfect accord with the simulation outcomes.\\
\end{abstract}


\section{Introduction}
In cognitive radio network, a secondary user (SU) is allowed to access the primary user (PU) spectrum completely if the available spectrum is not used by PU (interweave) or concurrently (underlay) with PU. The concurrent transmission is allowed if and only if the SU maintains a certain power threshold constraint known as interference temperature \cite{ekin2012random,aissacapacity}. Most of the studies \cite{ekin2012random,aissacapacity,PrimaryInpact,derivation,ekin2012capacity,PeakVsAverage} that involve this interference temperature model utilize either peak or average transmit power adaptation for the purpose of analyzing or improving the performance of secondary network. However, the impact of SU interference on primary network is let off completely. Therefore, to asses the performance and other quality of service (QoS) parameters, closed form expression need to be derived and validated. This study focuses on the mathematical foundation to derive these necessary performance expressions. This theoretical analysis is done by first considering interference from a  single SU and then extended to the case of interference from multiple SUs  on primary network. As per our knowledge, this is the first paper to analytically analyze the effect of SU interference on primary network.  The contribution of this paper can be summarized as follows:
\begin{itemize}
    \item Probability distribution function (PDF) and cumulative distribution function (CDF) expressions for  noise plus interference, signal to noise and interference ratio (SINR) are derived for both cases of interference from a  single SU and  multiple SUs on PU network.
    \item Closed form mean SINR expression, mean capacity and outage probability expressions are derived.
    \item The expression generated above are validated with simulation results to show the accuracy of the theoretical expressions.
\end{itemize}

The rest of the paper is organized as follows. In Section II, system model with underlying assumptions is described in detail. Section III presents the extensive theoretical analysis and comparisons with simulation results and finally, conclusions are given in Section IV.

\section{System Model }
\begin{figure}[!ht]
\centering
\includegraphics[width=\linewidth]{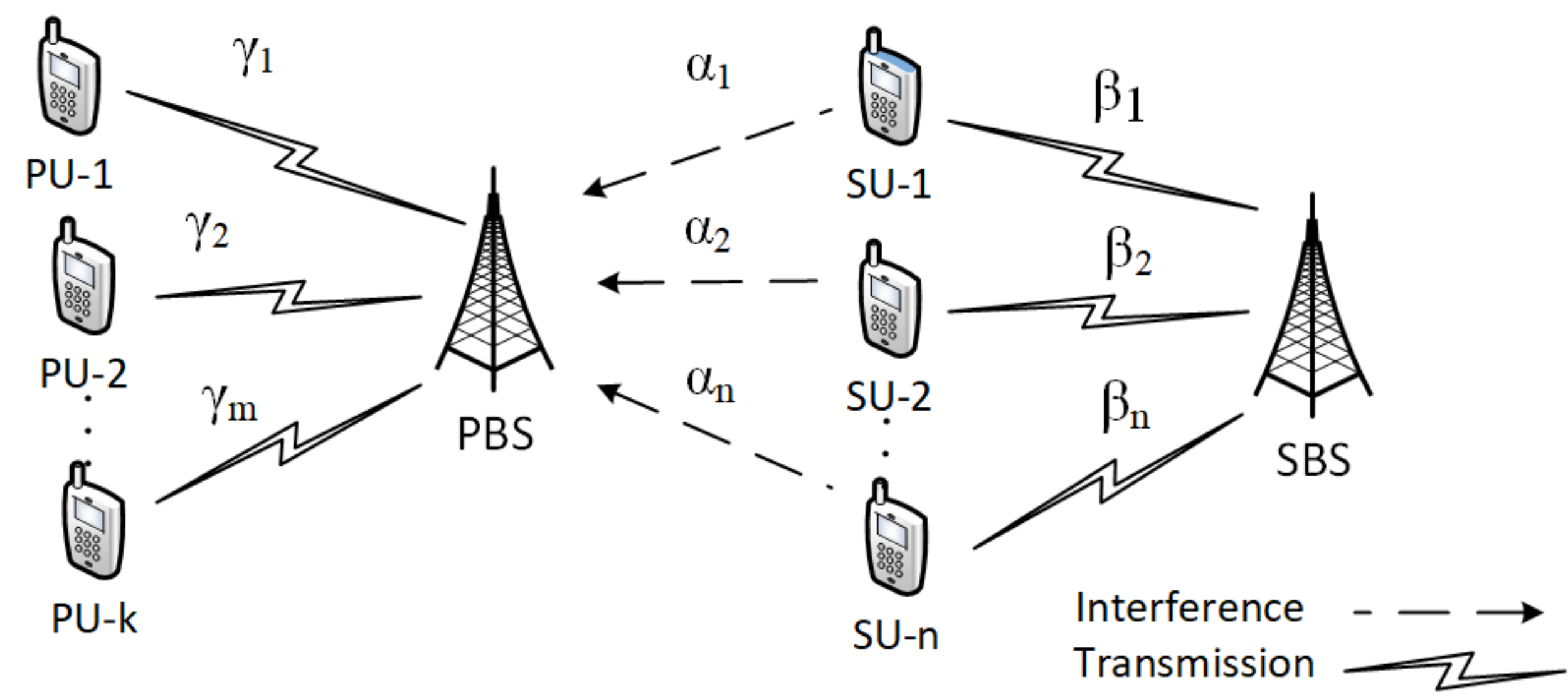}
 \caption{Underlay cognitive network with $n$-SUs sharing the spectrum with PU network of $k$-PUs. The  channel power gain between any SU-i ($i^{th}$ user) and PBS is denoted by $\alpha_i$, between any SU-i ($i^{th}$ user) and SBS by $\beta_i$ and between  $i^{th}$-PU and PBS by $\gamma_i$.}
\label{system}
\end{figure}

The cognitive radio network that is under consideration in this work is shown in Fig. \ref{system}. It consists of  $n$ SUs (SU network) and $k$ PUs with corresponding SBS and PBS. Since we are analyzing the effect of SU on PU network, we don't need to consider the channel gain between PU and  SBS, and also we don't need to consider the interference among PUs because of orthogonal resource allocation between them.

In addition, since the channel fading is assumed to be Rayleigh distributed, the channel power gains follow an exponential distribution. Considering the peak power adaptation  \cite{PeakVsAverage,derivation} for the case of interference from single SU on primary network, the secondary transmit power is given by:
\begin{equation}
P^{tx}=\text{min} \left \{ p , \frac{q}{\alpha} \right \}
\end{equation}

In the following sections, for theoretical analysis purpose, we assume that there are $n$ SUs that form the underlay cognitive network with primary user, where $n=\{1,2,3 \dots, n\}$. Furthermore, the thermal additive white Gaussian noise (AWGN) in the network is assumed to have circularly symmetric complex Gaussian distribution with zero mean and variance as $\sigma^2$, i.e., $\mathcal{CN}(0, \sigma^2)$.
\section{Theoretical Analysis}
In this section, theoretical expressions for PU performance parameters with interference  from a single SU will be derived first and then extended to the case of interference from multiple SUs. The interference observed at primary receiver  because of a single SU and multiple SUs with peak power adaptation will be given as:
\begin{equation} \label{I}
\begin{split}
   I_\text{single}&=\alpha P_{sec} =\text{min}   \{ \alpha p , q \},\\
    I_{\text{multi}}&=\text{min}  \left \{ \sum_{i=1}^{n} \alpha_i p , q \right \}. 
\end{split}
\end{equation}

Eq. \eqref{I} represents a  minimum of a random variable and a constant\footnote{\label{note1} For illustration purposes, the value's of peak power $p$ and interference temperature $q$ in this paper are chosen to be in linear scale. However, the expressions derived in this paper hold for any value of $p$ and $q$ for any scale.}. From the theory of mixed random variables\cite{papoulis2002probability,miller2012probability,pishro2016introduction}, a constant $c$  can be modelled as a random variable with PDF  equal to $\delta(x-c)$ and  CDF  equal to $H(x-c)$, where $H (x)$ is a Heaviside function and  $\delta (x)$ is a Dirac Delta function. So with interference from  a single SU, the CDF and PDF of  minimum of two independent random variables is then given by:
\begin{equation}{\label{cdfmethod}}
    \begin{split} 
     F_{I}(x)&=F_{\alpha p_{0}} (x)+F_{q}(x)-F_{\alpha p_{0}} (x)F_{q}(x), \\
     %F_{I}(x)&= (1-e^{-\frac{\lambda x}{p}})+ H(x-q)-(1-e^{-\frac{\lambda x}{p}}) H(x-q) \\
      F_{I}(x)&=1-e^{-\frac{\lambda x}{p}}(1- H(x-q)). 
      \end{split}
\end{equation}

On including noise $\mathcal{CN}(0, \sigma^2)$, the  CDF  of noise plus interference will be then,
\begin{equation}\label{CDF}
    F_{IN}(x)=1-e^{-\frac{\lambda (x-\sigma^2)}{p}}(1- H(x-\sigma^2-q)).
\end{equation}

Correspondingly, the PDF of interference and  noise is given by differentiating the CDF with respect to noise and interference variable  $x$, i.e.
\begin{equation} \label{PDF}
\begin{split}
       f_{IN}(x)&=\frac{\lambda}{p}e^{-\frac{\lambda (x-\sigma^2)}{p}} \Bigl( 1-H(x-\sigma^2-q)  \\
       &~+ \frac{p}{\lambda}\delta(x-\sigma^2-q) \Bigr ), \ \   \forall \  \sigma^2\le x\le \infty.
       \end{split}
       \end{equation}

For the case of interference from multiple SUs,  the distribution of of interference given in Eq. \eqref{I} will follow Gamma distribution\footnote{\label{note2} The distribution of sum of independent exponential random variables with the same rate parameters follows  Gamma distribution. Also, to distinguish between the channel $\gamma$  between PU and PBS, the Gamma distribution is  denoted  as $\bar{\gamma}$ in this study.},  $f_ {\bar{\gamma}} (x,\kappa,\theta)$, where $\kappa$ and $\theta$ represent the shape and rate parameter. The PDF and CDF of Gamma distribution is given as
\begin{equation} \label{gammadist}
\begin{split}
   f_{\bar{\gamma}}(x)&= \sum_{i=1}^{n} \alpha_i p = \bar{\gamma} \left(x,n,\frac{\lambda_2}{p}\right )= \bar{\gamma} \left(x,n,\bar{\lambda}\right ) ,\\
   &= \frac{\bar{\lambda}^{n} x^{n-1}}{\Gamma(n,0)}e^{-\bar{\lambda} x}, \ \forall \ \left \{x \ge 0 ,n >0 ,\bar{\lambda} >0 \right \},\\
   F_{\bar{\gamma}}(x)&=1- \frac{\Gamma \left(n,\bar{\lambda}x \right)}{\Gamma(n,0)}, \ \forall \ \left \{x \ge 0 ,n >0 ,\bar{\lambda} >0 \right \},
\end{split}
  \end{equation}
where  $n$ is the total number of SUs in the underlay network, $\bar{\lambda}= \lambda_{2}/p$ is the scaled rate parameter between  SU and PBS  and  $\Gamma(a,x)$ is an incomplete gamma function defined as:
\begin{equation}
    \Gamma(a,x)=\int_{a}^{\infty} t^{a-1}e^{-t} dt,  \ \forall \ a>0,x \ge 0. \nonumber
\end{equation}

\begin{figure}[!ht]
\centering
\includegraphics[width=\linewidth]{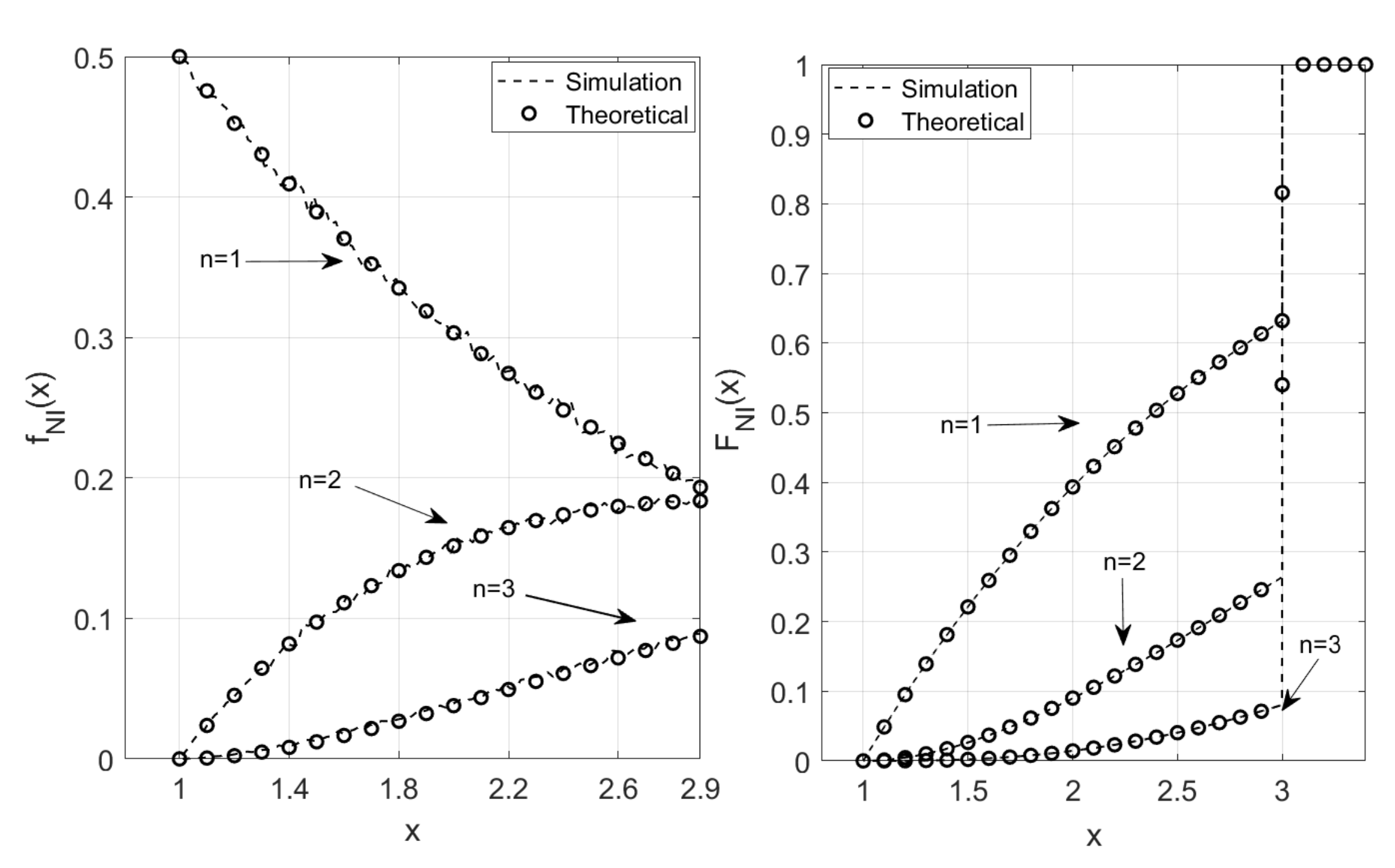}
 \caption{PDF and CDF of noise and interference for different number of SUs ($n=1,2,3$), when $p>q$, where $p=4$, $q=2$ and $ \sigma^2=1$ with support region from $ \sigma^2 \le x \le \infty$.}
\label{PDFCDFNI}
\end{figure}

\begin{figure}[!ht]
\centering
\includegraphics[width=\linewidth]{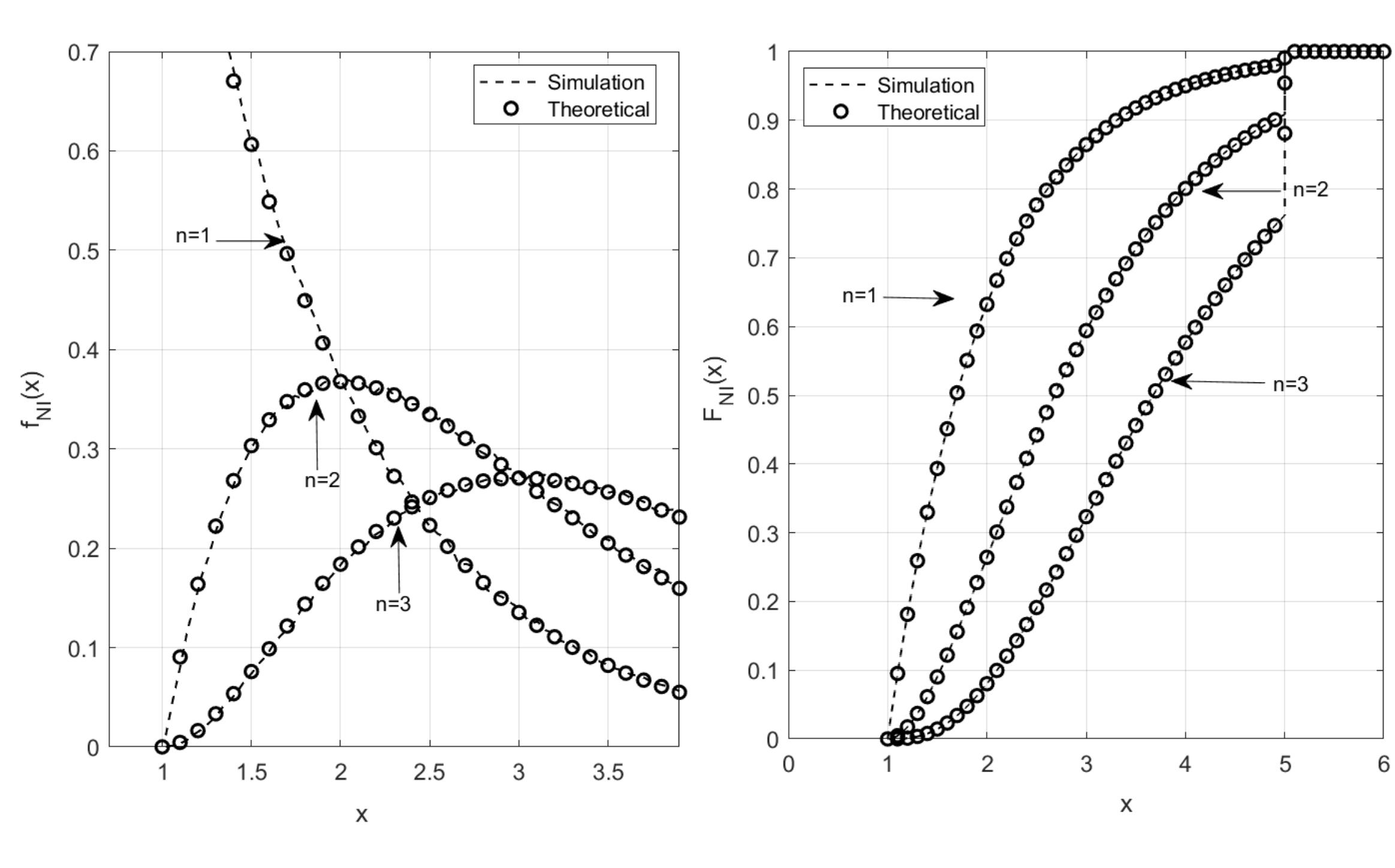}
 \caption{PDF and CDF of noise and interference for different number of SUs ($n=1,2,3$) when  $p<q$, where $p=2$  $q=4$ and $\sigma^2=1$ with support region from $ \sigma^2 \le x \le \infty$.}
\label{PDFCDFNI2}
\end{figure}

Following the same mathematical approach that was used in single SU case, the distribution of noise plus interference in multiple SUs case is then derived as:
\begin{equation}\label{NI_n}
    \begin{split} 
     F^{m}_{NI}(x)&=1- \frac{\Gamma \left(n,\bar{\lambda} (x-\sigma^2) \right)}{\Gamma(n,0)} + H(x-\sigma^2-q) \\
     &~~~~~~~~~~~~~~~~~~~~~~~~~~~~~ \times \frac{\Gamma \left(n,\bar{\lambda}(x-\sigma^2) \right)}{\Gamma(n,0)} ,\\
     f_{NI}^{m}(x)&=  \frac{\Gamma \left(n,\bar{\lambda} (x-\sigma^2) \right)}{\Gamma(n,0)}  \delta(x-\sigma^2-q)+ \bar{\lambda}^n    \\
     &\times\frac{(x-\sigma^2)^{n-1}}{\Gamma(n,0)}  e^{-\bar{\lambda} (x-\sigma^2)} [1-  H(x-\sigma^2-q)],
\end{split}
\end{equation}
where $\sigma^2$ is the $\mathcal{CN}(0, \sigma^2)$. Fig. \ref{PDFCDFNI} and Fig. \ref{PDFCDFNI2} plots the  CDF and PDF for the theoretical expression (Eq. \eqref{NI_n} )  with the simulation result of  $p>q$ and  $q<p$ for different SU densities of $n=1,2,3$.
\subsection{Instantaneous SINR}
The instantaneous SINR at PBS considering the system model (Fig. \ref{system}) is given by:
\begin{equation}
    \text{SINR}=\frac{\gamma p}{\sigma^2+I},
\end{equation}
where $I$ is the interference from SUs given by Eq. \eqref{I}. The distribution of numerator is a scaled exponential distribution and the distribution of denominator is already derived in the previous section (Eq. \eqref{PDF} and Eq. \eqref{NI_n}). Therefore, the PDF of ratio of two  independent random variables \cite{curtiss1941} i.e. $z=x/y$, where $x=\gamma p$ and $y=\sigma^2+I$ will be  given  as
\begin{equation} \label{ratio}
\begin{split} 
f_{z}(z)& = \int_{\sigma^2}^\infty  y \cdot f_{x,y}(yz,y) dy= \int_{\sigma^2}^\infty y \cdot f_x(yz)f_{y}(y)dy \\
 & ~~~~~~~~~~~~~~~~~~~~~~~~~~~~~~~~~~~~~~~~~~~~~\forall \ y \ge 0.
\end{split}
\end{equation}

For the interference from a single SU user, the SINR distribution will be as follows:
\begin{equation}\label{SINR_single}
\begin{split} 
f_{z}(z) & =\int_{\sigma^2}^\infty y\cdot \frac{\lambda_{1} e^{\frac{-\lambda_{1} yz}{p}}}{p}  \frac{\lambda_2}{p}e^{-\frac{\lambda_2 (y-\sigma^2)}{p}}, 
   \nonumber  \\
&\times \Bigl \{  1-H(y-\sigma^2-q) +\frac{p}{\lambda_2}\delta(y-\sigma^2-q) \Bigr\} dy \nonumber\\
 & =\frac{\lambda_1 \lambda_2}{p^2} e^{\frac{\lambda_2 \sigma^2}{p}} \left \{     \int_{\sigma^2}^\infty y \cdot e^{\frac{-y(\lambda_1 z+ \lambda_2)}{p}} dy   -  \int_{\sigma^2+q}^\infty y  \right. \nonumber\\
 & \times e^{\frac{-y(\lambda_1 z+ \lambda_2)}{p}}  dy \left. +  \frac{p}{\lambda_2} (\sigma^2+q)  e^{\frac{-(\sigma^2+q)(\lambda_1 z+ \lambda_2)}{p}} \right \} . \nonumber
\end{split}
\end{equation}

By using integration by parts and on further simplification, the PDF is  reduced to:
\begin{equation}
    \begin{split}
        f_{z}(z) &=\frac{\lambda_1 \lambda_2}{\Lambda p} \left \{ e^{\frac{-\sigma^2 \lambda_1 z}{p}} \left( \sigma^2+ \frac{p}{\Lambda} \right)\right. \nonumber \\
     & \left. +\  e^{-\frac{\sigma^2  \lambda_1 z +q \Lambda}{p}} \left(\frac{(\sigma^2+q)\lambda_1 z}{\lambda_2}- \frac{p}{\Lambda} \right)\right \}, 
    \end{split}
\end{equation}
where $\Lambda$  is the scaled and shifted random variable version \footnote{\label{note3}$\lambda_1$ is the channel rate parameter between PU and PBS, whereas $\lambda_2$ is the channel rate parameter between SU and PBS.} of $z$ given by $\Lambda=\lambda_1+ \lambda_2 z $ . 
Under the scenario of $\lambda_1 = \lambda_2= 1$, with AWGN as $\mathcal{CN}(0, \sigma^2=1)$, the PDF can be further simplified to
\begin{equation} \label{PDFSnr}
\begin{split}
   f_{z}(z)& =\frac{1}{p (z+ 1)} \left \{ e^{\frac{- z}{p}} \left( 1 + \frac{p}{ z+ 1}  \right)  + e^{-\frac{ z +q (  z+ 1)}{p}} \right.\\
      &~~~~~~~~~~~~~~~~~~~~~~~ \left. \times \left((1+q) z- \frac{p}{ z+ 1} \right)\right \}.  
\end{split}
       \end{equation}

\begin{figure}[!ht] 
\centering
\includegraphics[width=\linewidth]{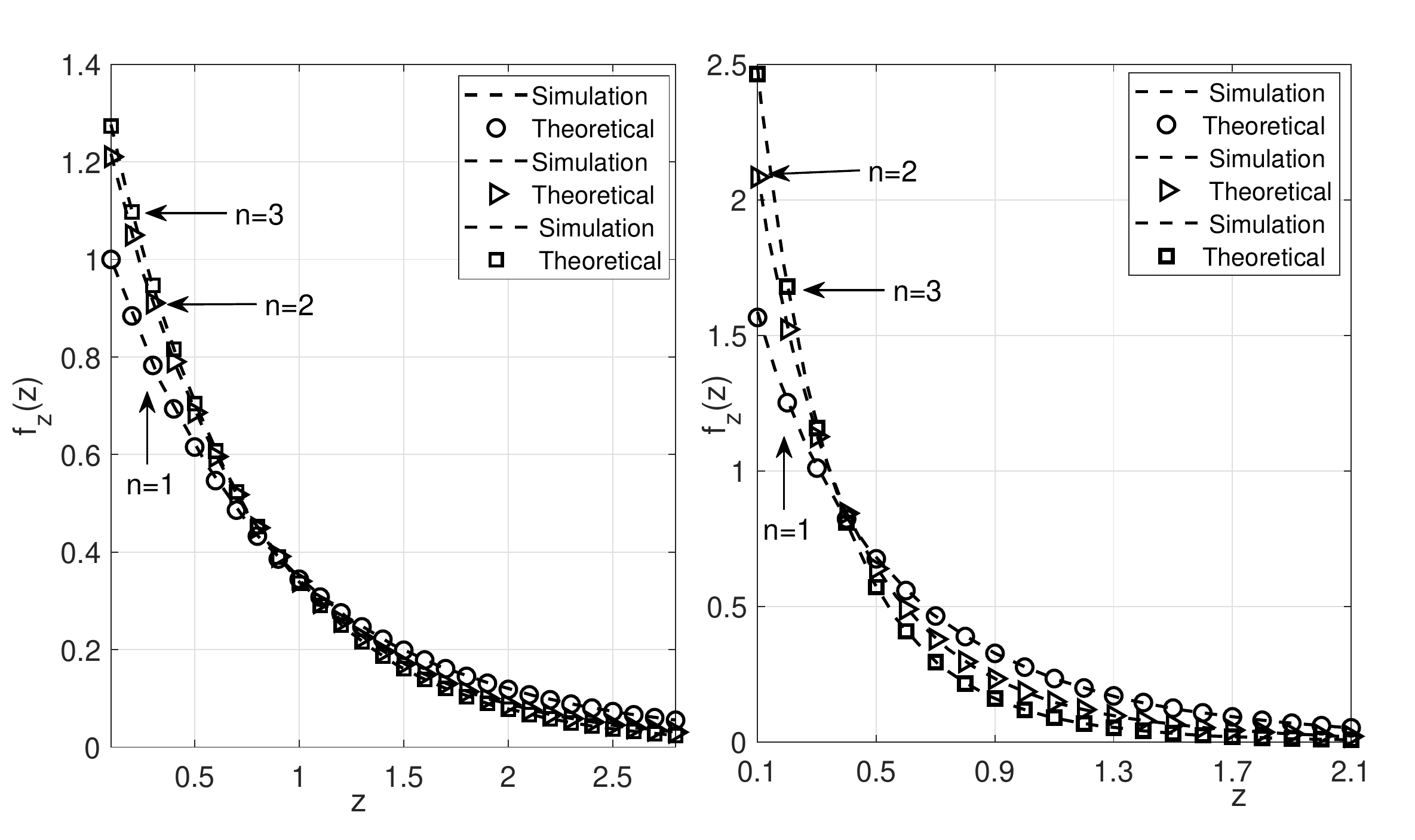}
 \caption{PDF of SINR for two cases of  $p<q$ and  $p>q$ for different number of SUs ($n=1,2,3$).}
\label{PDFcSINR}
\end{figure}

Following the same analytical framework used for the single SU case, the distribution with interference from multiple SUs will be:
\begin{equation} 
\begin{split} 
f_{z}^{m}(z)&=\int_{\sigma^2}^\infty y \bar{\lambda_{1}} e^{-\bar {\lambda_{1}}yz} \Big \{  f_{\bar{\gamma}} (y-\sigma^2,n,\bar{\lambda}) + \delta(y-\sigma^2-q)  \\
& ~~~~~~~~~~~ \times (1-F_{\bar{\gamma}} (y-\sigma^2,n,\bar{\lambda}))- f_{\bar{\gamma}} (y-\sigma^2,n,\bar{\lambda})\\
&~~~~~~~~~~~~~~~~~~~~~~~\times H(y-\sigma^2-q) \Big \} dy. \nonumber\\
\end{split}
\end{equation}

which on further evaluation and simplification reduces to
\begin{equation} \label{multiSUsinr}
\begin{split}
f_{z}^{m}(z)&=\bar{\lambda_1}  \bar{\lambda}^{n} e^{-\sigma^2 \bar{\lambda_1}z} \Theta^{-1-n} 
\Bigg [ n + \bigg ( \sigma^2 \Theta\\
& \times \bigg [ 1- \frac {\Gamma(n,q\Theta)}{ \Gamma(n,0)} \bigg ] \bigg ) - \frac{ \Gamma(n+1,q\Theta)}{\Gamma(n,0)}\Bigg ]  \\
&~~~~~~~~ + \bar{\lambda_1} (\sigma^2+q) \frac{\Gamma(n,q \bar{\lambda})}{\Gamma(n,0)} e^{-\bar {\lambda_{1}}(\sigma^2+q) z },
\end{split}
\end{equation}
where $\Theta$ is the scaled and shifted random variable version \footnote{\label{note4} Here $\bar{\lambda}= \lambda_{2}/p$ is the scaled rate parameter of SUs and  $\bar{\lambda_1}=\lambda_{1}/p$ is the scaled rate parameter for PU.} of $z$ given by $\Theta=\bar{\lambda}+ \bar{\lambda_1}z $. Fig. \ref{PDFcSINR} shows the plot of the derived theoretical expression with simulation data for the two cases of  $p<q$ and $p>q$ with different SU densities ($n=1,2,3$).

In the following sections, we will look into the crucial performance metrics (outage probability and capacity) of underlay cognitive network. The mentioned approach can be extended to the case of interference from multiple SUs given that the important SINR expression  Eq. \eqref{multiSUsinr} for multiple SUs is already been derived. However, given the space limitations, the derivations considering multiple SUs are not detailed herein in coming sections. Nonetheless, the fundamental case of interference from single SU case has been presented in detail.

\subsection{Mean SINR}

The mean SINR is given as $\mu=\int_{0}^\infty  z f_{z}(z) dz $ where the PDF of SINR $f_{z}(z)$ was derived in  Eq. \eqref{PDFSnr}. Thus,
\begin{equation}
    \begin{split} 
      \mu &= \int_{0}^\infty  z \Bigl \{ \frac{1}{p (z+ 1)} \Bigl\{ e^{\frac{- z}{p}} \left( 1 + \frac{p}{ z+ 1} \right)   \nonumber\\
      & ~~~ + e^{-\left(\frac{ z +q (  z+ 1)}{p}\right)} \left((1+q) z- \frac{p}{ z+ 1} \right)\Bigr\}  dz, \nonumber \\
      &=\int_{0}^\infty \left ( \frac{ze^{\frac{- z}{p}}}{p (z+ 1)} \right) dz +  \int_{0}^\infty \left ( \frac{ze^{\frac{- z}{p}}}{ (z+ 1)^2} \right) \ dz  \\
    &~~~~~~~~+  \int_{0}^\infty \left( \frac{z^2 (1+q)e^{-\left(\frac{ z +q (  z+ 1)}{p}\right)}  }{p (z+1)}  \right) dz  \nonumber  \\
        &~~~~~~~~~~~~~~ - \int_{0}^\infty \left( \frac{z e^{-\left(\frac{ z +q (  z+ 1)}{p}\right)} }{ (z+1)^2}                  \right) dz .\nonumber\\
         \end{split}
     \end{equation}

which on further simplification reduces to
     \begin{equation}
         \begin{split}
              \mu  &= e^{\frac{1}{p}} \left \{ \Gamma{\left(0,\frac{1}{p}\right)} - \Gamma{\left(0,\frac{1+q}{p}\right)} \right \} + \frac{p e^{\frac{-q}{p}}}{1+q}.
         \end{split}
     \end{equation}
\begin{figure}[!ht] 
\centering
\includegraphics[width=\linewidth,height=5.5cm]{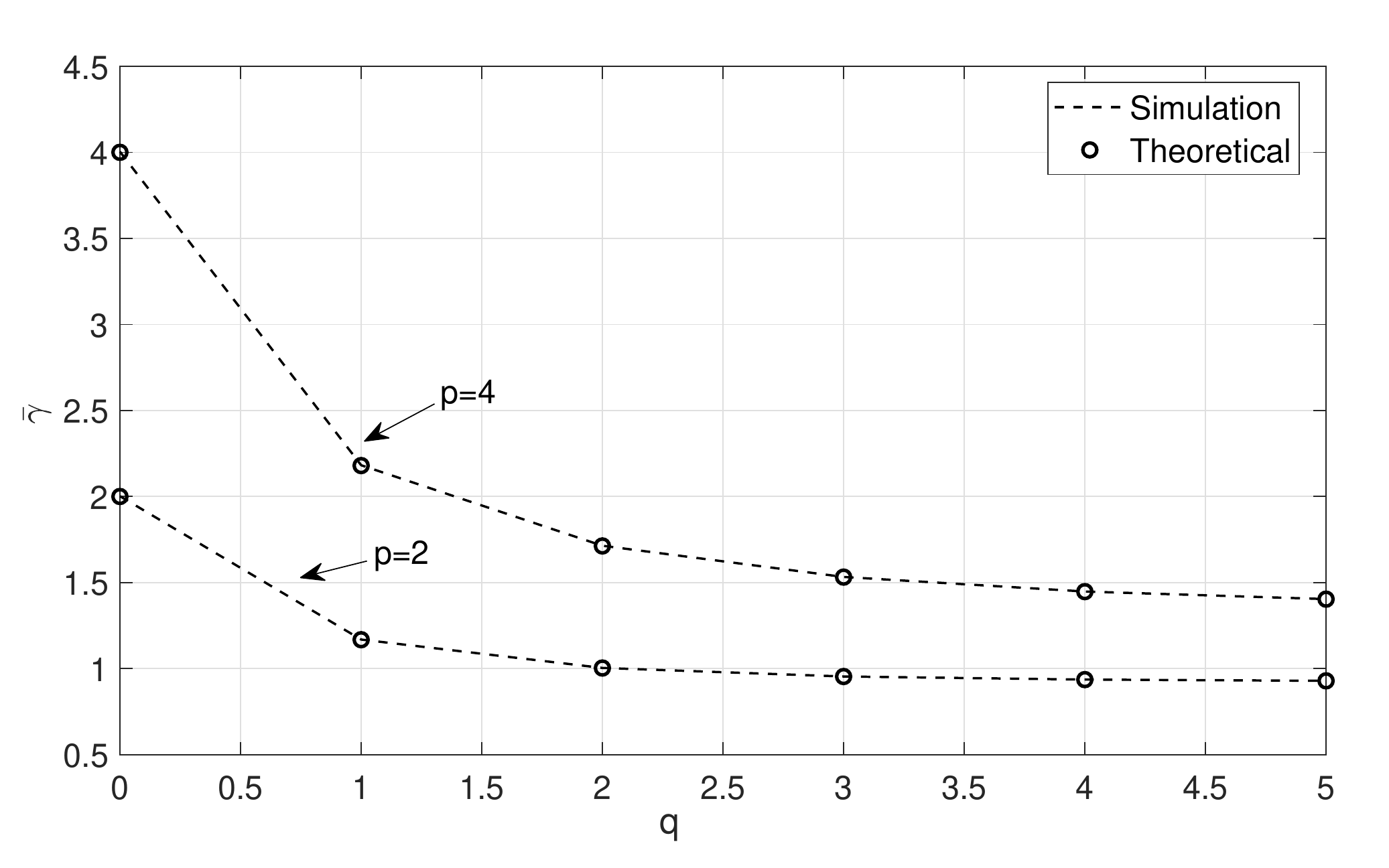}
 \caption{Mean SINR vs Interference Temperature, $q$, for $p=2$ and $p=4$.}
\label{meanSINR}
\end{figure}
Fig. \ref{meanSINR} shows that the change in mean SINR while varying the interference temperature $q$ for a constant peak transmit power $p$. The  higher transmit power (for both PU and SU) with lower interference temperature gives better mean SINR than lower transmit power (for both PU and SU) with high IT constraint. 

\subsection{Outage Probability of Primary Network}
The outage probability  is defined as the probability when the instantaneous SINR drops below a given threshold. Mathematically, this is given as: $Pr(\gamma \le \psi)=F_{z}(\psi)$, which is nothing but the CDF of SINR. Therefore,
\begin{equation}
   \begin{split}
     F_{z}(\psi)& = \int_{0}^\psi f_{z}(z) dz, \nonumber \\
          &=\int_{0}^\psi \left ( \frac{e^{\frac{- z}{p}}}{p (z+ 1)} \right) dz +  \int_{0}^\psi \left ( \frac{e^{\frac{- z}{p}}}{ (z+ 1)^2} \right)dz \nonumber \\
       & ~~~~+ \int_{0}^\psi \left( \frac{z (1+q)e^{-\left(\frac{ z +q (  z+ 1)}{p}\right)}  }{p (z+1)}  \right) dz  \nonumber  \\
        & ~~~~~~~~~  - \int_{0}^\psi \left( \frac{ e^{-\left(\frac{ z +q (  z+ 1)}{p}\right)} }{ (z+1)^2}  \right) dz .\nonumber
   \end{split}
\end{equation}

which on  further integration and simplification reduces to
\begin{equation}
    \begin{split}
      F_{z}(\psi)& =1- \frac{e^{\frac{- \psi}{p}}}{\psi+1} \left( 1+ \psi e^{\frac{- q (\psi +1)}{p}} \right ) \nonumber 
             \end{split}
\end{equation}
\begin{figure}[!ht]
\centering
\includegraphics[width= \linewidth,height=5.5cm]{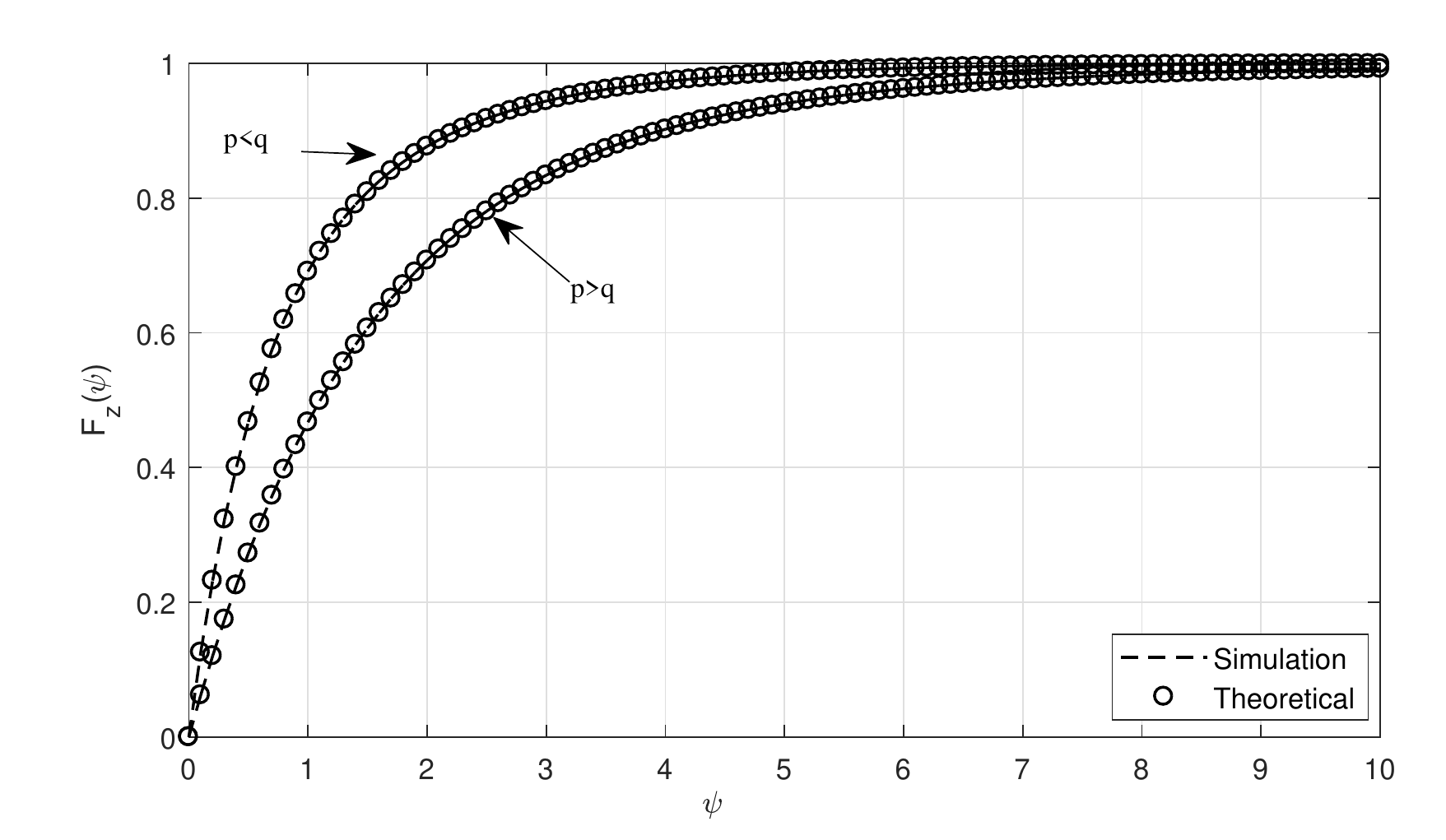}
 \caption{Outage probability of PU for $p<q$, where $p=2$ and $q=4$ and for $p>q$, where $p=4$ and $q=2$. } 
\label{CDFout}
\end{figure}

It can be directly inferred from Fig. \ref{CDFout} that if $q>p$,  the outage probability is higher than in the case of $p>q$. In addition to this inference, it can be also observed that the theoretical expressions derived are in sync with the simulation results, i.e., increase the spectral efficiency of the network. 
\subsection{Instantaneous Capacity of Primary Network}
The PDF of instantaneous capacity can be readily found from the PDF of instantaneous SINR by using transformation of random variables method \cite{papoulis2002probability,miller2012probability}. This can be obtained by using:
\begin{eqnarray}
      f_{x}(x)=f_{z}(z)\Bigl |\frac{d z}{dx} \Bigr |_{z=e^x -1},\nonumber
\end{eqnarray}
 where $f_{z}(z)$ is derived in Eq. \eqref{PDFSnr} for the case of interference from single SU on primary network.
 \begin{figure}[!ht]
\centering
\includegraphics[width=\linewidth,height=5.5cm]{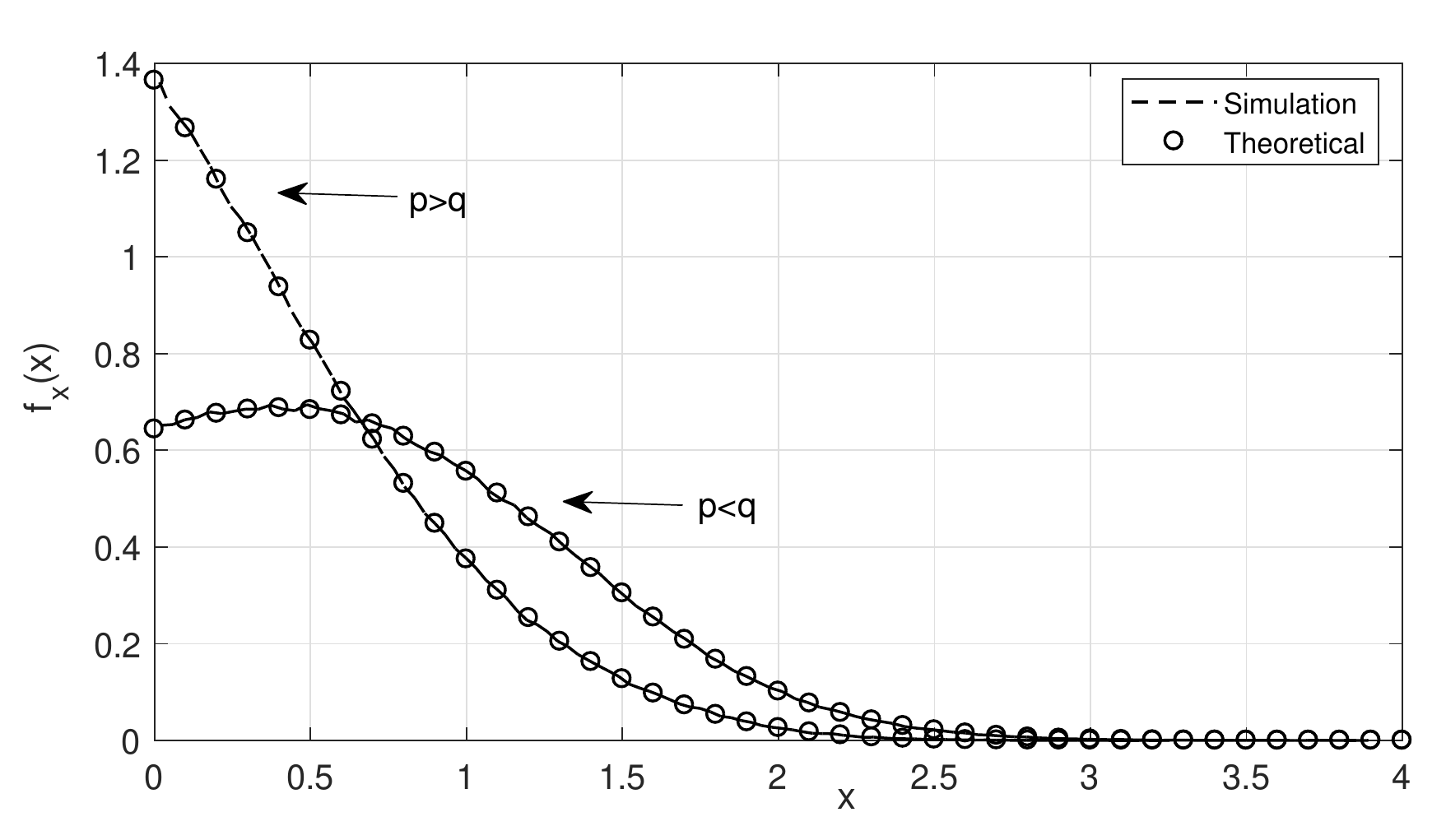}
 \caption{Instantaneous PDF of Capacity  for  $p<q$, where $p=2$ and $q=4$ and for $p>q$, where $p=4$ and $q=2$.}
\label{PDFcap}
\end{figure}
It can be seen from  Fig. \ref{PDFcap} that there is a point where the instantaneous capacity for $p<q$ goes below $p>q$. It proves the point that the interference temperature should not be kept constant rather should be dynamic in nature to exploit full potential of the network.
 \subsection{Mean Capacity}
The average capacity from the PDF of instantaneous SINR ($f_{z}(z)$)  is given as:
 \begin{eqnarray}
       \bar{C}=\int_{0}^{\infty}  \text{log}(1+ z)f_{z}(z) dz. \nonumber
 \end{eqnarray}

Substituting Eq. \eqref{PDFSnr} in the above expression and on further evaluation.
 \begin{equation} \label{Capmean}
    \begin{split}  
       \bar{C} &=  \frac{e^{\frac{1}{p}}}{p} \left [  \Gamma\left(0,\frac{z+1}{p} \right) -  \Gamma\left(0,\frac{(q+1)(z+1)}{p}\right) \right.  \nonumber  \\
          & ~~~~~~~~ \times (p+q+1) \bigg ]  -\frac{e^\frac{1-(q+1)(z+1)}{p}}{z+1} \left \{ e^\frac{q(z+1)}{p}   \right. \nonumber \\
         &~~~~~~~~~~~\times \Big ( 1 +\text{log}(z+1) \Big ) +   z\text{log}(z+1) -1  \Big \}  \Bigg|_0^\infty .
    \end{split}
\end{equation}
\newline

At $z=\infty$, $\Gamma(0,z) \to 0 $ and also, $\frac{e^\frac{1-(q+1)(z+1)}{p}}{z+1} \to 0$.\\
Therefore,  the final mean capacity expression will be evaluated at $z=0$:
\begin{equation} \label{FinCapmean}
    \begin{split}  
      \bar{C}& =1-e^{-\frac{q}{p}}+ \frac{e^{\frac{1}{p}}}{p} \Bigg [ (p+q+1)  \Gamma\left(0,\frac{q+1}{p}\right)- \Gamma\left(0,\frac{1}{p} \right) \Bigg ]\nonumber
    \end{split} 
    \end{equation}

\begin{figure}[!ht]
\centering
\includegraphics[width=\linewidth,height=5.6cm]{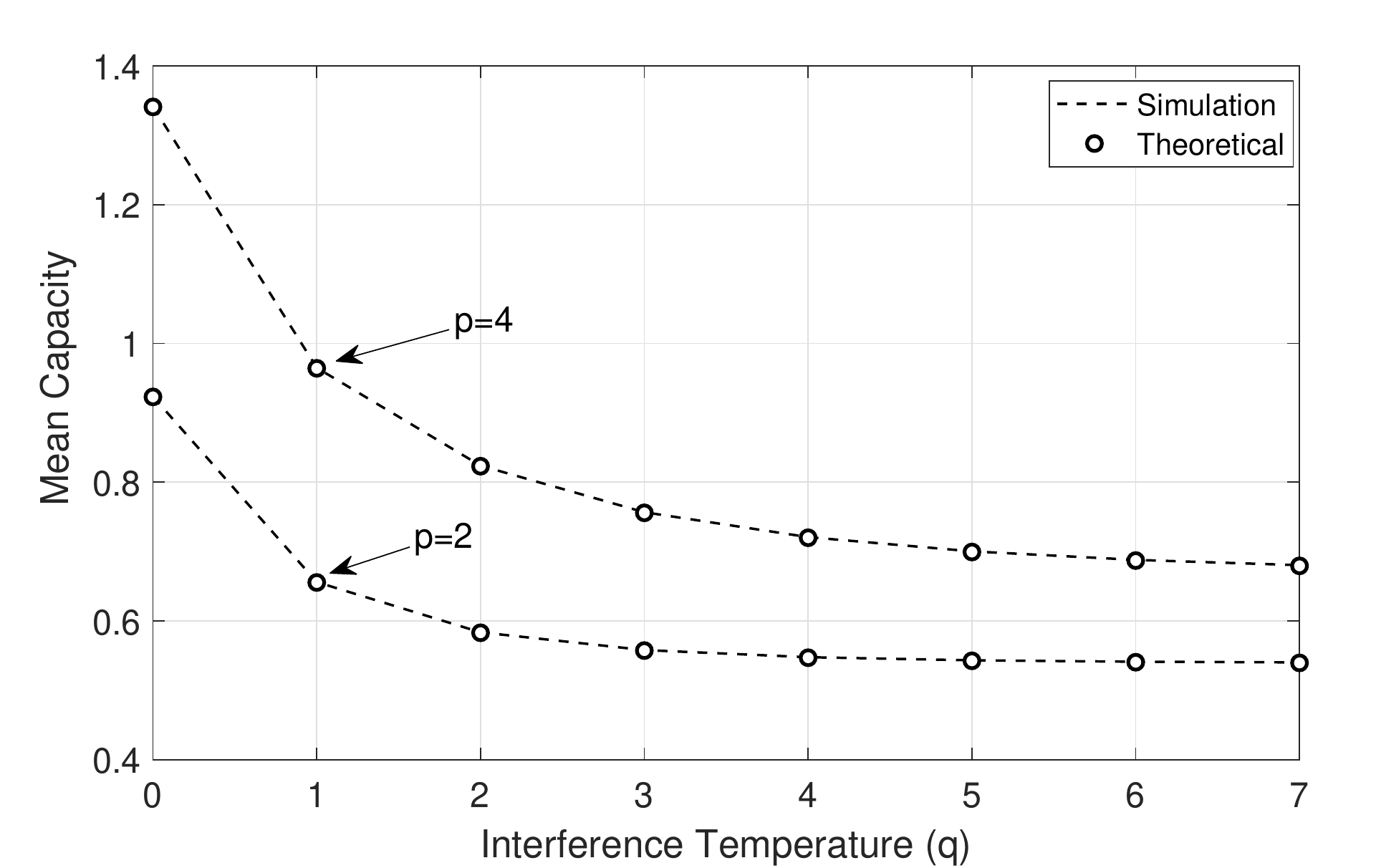}
 \caption{Theoretical and simulation result plots for  capacity  at $p=2$ and $p=4$ with varying interference temperature: $q$.}
\label{Meancap}
\end{figure}

Fig. \ref{Meancap} shows the plot of this theoretical expression with simulation results for two cases of transmit power $p=2$ and $p=4$. Intuitively, high transmit power $p=4$ will result in high capacity for the network  than the low transmit power of $p=2$ but when the interference temperature is relaxed, the interference caused due to secondary user on primary will also increase that in turn will reduce the overall capacity.
\section{Conclusion}
In this paper, the performance of primary network is studied considering interference from the SU network. The analysis is done under peak power adaptation method at secondary transmitter. Given the importance of dynamic interference temperature for network performance, closed form expressions for the PDF and CDF of interference and noise, SINR for interference from single and  multiple SUs are derived. Furthermore, instantaneous capacity  with theoretical  expressions for mean SINR, mean capacity and   outage probability are deduced for simplistic network consisting of interference from single SU. Finally, the theoretical expressions are validated with the simulation results.
\section*{Acknowledgment}
This work was supported by NASA Oklahoma Space Grant Consortium (EPSCoR) Research Initiation Grant. Also, the authors would like to thank the TPC and reviewers for their valuable feedback and suggestions.

% conference papers do not normally have an appendix

% use section* for acknowledgment

% trigger a \newpage just before the given reference
% number - used to balance the columns on the last page
% adjust value as needed - may need to be readjusted if
% the document is modified later
%\IEEEtriggeratref{8}
% The "triggered" command can be changed if desired:
%\IEEEtriggercmd{\enlargethispage{-5in}}

% references section

% can use a bibliography generated by BibTeX as a .bbl file
% BibTeX documentation can be easily obtained at:
% http://mirror.ctan.org/biblio/bibtex/contrib/doc/
% The IEEEtran BibTeX style support page is at:
% http://www.michaelshell.org/tex/ieeetran/bibtex/
%\bibliographystyle{IEEEtran}
% argument is your BibTeX string definitions and bibliography database(s)
%\bibliography{IEEEabrv,../bib/paper}
%
% <OR> manually copy in the resultant .bbl file
% set second argument of \begin to the number of references
% (used to reserve space for the reference number labels box)

\bibliographystyle{IEEEtran}

\bibliography{biblog}

% that's all folks
\end{document}